# Fast formation and superconductivity of $MgB_2$ thick films grown on stainless steel substrate

A.H. Li, X.L. Wang, M. Ionescu, S. Soltonian, J. Horvat, T. Silver, H.K. Liu, and S.X. Dou

Institute for Superconducting and Electronic Materials, University of Wollongong, NSW 2522, Australia

## Abstract

The fabrication, characterisation, and superconductivity of $MgB_2$ thick films grown on stainless steel substrate were studied. XRD, SEM, and magnetic measurements were carried out. It was found that the $MgB_2$ thick films can be fast formed by heating samples to 660 $^{o}$C then immediately cooling down to room temperature. XRD shows above 90% $MgB_2$ phase and less than 10 % MgO. However, the samples sintered at 800 $^{o}$C for 4 h contain both $MgB_4$ and MgO impurities in addition to $MgB_2$. The fast formed $MgB_2$ films appear to have a good grain connectivity that gives a $J_c$ of 8 x 10 $^4$ A/cm$^2$at 5 K and 1 T and maintained this value at 20 K in zero field.

## Introduction

The discovery of superconductivity at 39 K in $MgB_2$ [1] has generated great interest worldwide in both fundamental studies and practical applications worldwide. Critical current densities on the order of $10^{4}$ to $10^{5}$ A/cm$^2$ have been reported by several groups for polycrystalline $MgB_2$ bulk samples [2-8]. A strong link type critical current density has been observed, regardless of the degree of grain alignment [3]. This would be an advantage for making wires or tapes with no degradation of $J_c$, unlike the degradation due to the grain boundary induced weak-links which is a common and a serious problem widely encountered in cuprate high temperature superconductors. $H_{c2}$ of $MgB_2$ has been determined to be as high as 14 T at 4.2 K. A significantly large $J_c$ of $10^{6}$ A/cm$^2$ at 4.2 K and 1 T and enhancement of the irreversibility line have been reported in high quality epitaxial $MgB_2$ thin films grown on $Al_2O_3$ and $SrTiO_3$ single crystal substrates [9, 10]. This result gives further encouragement of to the development of $MgB_2$ for high current applications. Fabricating $MgB_2$ into tapes or wires will be essential for most such applicants. The first short wire of $MgB_2$ reported was made by exposing boron filament to magnesium vapour [11], with a $J_c$ of $10^5$ A/cm$^2$. For tape fabrication, it is very important to find a suitable sheath material for $MgB_2$ which does not degrade the superconductivity. The first tape of $MgB_2$ reported was made using Nb as sheath material [12]. The motivation for our study on the fabrication of $MgB_2$ thick films on different substrates is to find a suitable sheath material for making $MgB_2$ tapes, as thick film preparation is very much easier than making tapes. In addition, the thick films of $MgB_2$ have a great potential for applications in magnetic shielding devices and high power microwave devices that will work at significantly higher temperatures than conventional low $T_c$ superconductors, due to the high $T_c$ of 39 K. It was reported very recently that the surface resistance of a dense $MgB_2$ sample is lower than that of polycrystalline Y-123 bulks and matches with Y123 single crystals [13]. These results show promise for the use of $MgB_2$ in microwave applications. We have noted that there are plenty of elements in the periodic table that can form the $AlB_2$ structure, and that most of them have a high solubility in $MgB_2$ [14]. Because the supercurrent is most likely carried by holes, any electron doping reduce the hole concentration level and in turn degrades the $T_c$ of $MgB_2$ [15]. It has been shown that doping can degrade the $T_c$ in $Mg_{1-x}Al_xB_2$ [16]. It is noted that the size of iron is very small compared with Mg and therefore the ion has a very small solubility in $MgB_2$. In this letter, we report our study on the formation and properties of $MgB_2$ thick films grown on stainless steel substrate.

## Experimental

For the film preparation, a suspension of mixed magnesium and amorphous boron powder was made by stirring the powders together with acetone. The powders of Mg and B were deposited suspension after evaporation of acetone on a piece of polished stainless steel which placed at the bottom. This procedure was repeated several times until the desired thickness was reached. The resulting samples were pressed under high pressure after the deposition process in order to increase the density of the deposited films. Another piece of stainless steel of the same size as the substrate was placed onto the top of the deposited films so as to avoid the loss and oxidation of Mg during the sintering at high temperatures.

The samples were sintered in a tube furnace using two different temperature profiles (a) and (b) as shown in Fig. 1. For (a), the temperature was increased at a heating rate of 300 $^oC/h$ to 660 $^oC$, which is the exact melting point of magnesium, then furnace cooled down to room temperature without any holding period at 660 $^oC$. For (b), temperature was increased at the same heating rate of 300 $^{o}$C/h to 800 $^{o}$C, and held there for 4 h, then furnace cooled down to room temperature. A flow of high purity Ar gas was maintained throughout the whole sintering process.

The phases and film morphologies were investigated using X-ray diffraction (XRD) and scanning electron microscopy (SEM). Superconductivity was characterised using PPMS. The superconducting transition temperature was determined by measuring ac susceptibility as a function of temperature. Magnetisation hysteresis loops were recorded at 5 K and 20 K in fields up to 8 Tesla.

## Results and discussion

After removing the top piece of stainless steel, the films were found to be of a shinning with black colour. However, the films appear not to be firmly attached to the substrate. A plate-like layer of film could be easily removed from the substrate, probably indicative a non-reaction between the films and the stainless steel. This can be understood by considering that iron has a very small solubility in $MgB_2$ due to the fact that the iron is too small compared with the Mg. Another reason for the poor adherence of the film to the substrate is the formation of fine particles of MgO that may be acting as a barrier. This is evidenced by the SEM result as will be seen below. It is also possible that there may be a large difference in thermal expansion at high temperatures for steel and $MgB_2$ so that they separate from each other during cooling.
The thicknesses of the films are about 160 µm. The density of both samples is about 1.6g/cm$^3$.

Fig.2 shows the XRD patterns for samples sintered by temperature profile (a) (sample A) and (b) (sample B). It can be seen that most of the peaks can be indexed using the lattice parameters of $MgB_2$. Sample A only contains MgO impurities. However, sample B which was sintered at 800 $^{o}$C for 4 h, contains both MgO and $MgB_4$. This means that heating up to 660 $^{o}$C without further holding time is sufficient for $MgB_2$ phase formation for the full amount of materials used in the films. According to the Mg-B phase diagram, MgB2 can decompose to MgB4 only at high temperatures above 1545 $^{o}$C [17]. The sintering temperature of 800 $^{o}$C used for sample B is much lower than the decomposition temperature of MgB2, but it is much higher than the melting point of magnesium. Therefore, it is very likely that some of the magnesium evaporated at 800 $^{o}$C through the flow of Ar gas at the early stage of the reaction between magnesium and boron. This loss of magnesium at 800 $^{o}$C would have caused a change of the stoichiometry from the starting composition to the magnesium poor or boron rich side, which in turn resulted in the formation of MgB4 in addition to MgB2 as indicated in the phase diagram [14][17]. The formation of MgO may come from the starting Mg which may contain MgO.

Fig. 3 shows the SEM images of the two samples. Sample A shows a flat and dense surface (Figs 3(a) and (b)) with grain sizes of less than 1 μm. Sample B (fig. 3 (c)) had a rougher surface but similar grain size. The cracks seen in Fig. 3 (a) mainly occurred after handling. Fig. 3 (d) is the surface of the substrate after the $MgB_2$ film was removed. The grains from substrate are clearly seen with some small white MgO particles on top of them. The formation of MgO may be a barrier to the adherence of film to the substrate.

The $T_c$ of 37.5 K was determined for both samples by using a.c. susceptibility in an ac field of 1 Oe and a frequency of 117 Hz, as shown in Fig. 4. Both samples have a similar value of susceptibility. M-H loops measured at 5 and 20 K for both samples are shown in Fig. 5. A flux jump for H < 1 T can be seen for both samples at 5 K which is the same as what has been reported for a $MgB_2$ pellet with grain sizes as large as 200 μm [18]. In addition, the magnetisation hysteresis is much larger than for a weak-linked $MgB_2$ sample which was sintered at 800 $^oC$ using high pressure and reacted $MgB_2$ powders used as starting material [19]. This means that the grains are well connected, similar to what is formed in compacted $MgB_2$ pellets with large grain sizes of about 200 μm [18]. The critical current density can be obtained from the loops using the formula $J_c=20\ M/a(1-a/3b)$ ($a<b$) [20]. The $J_c$ as a function of temperature is shown in Fig. 6. The $J_c$ at 5 K for H < 1 T cannot be defined due to the flux jump. The $J_c$ of both samples is about 8 x $10^4$ A/cm$^2$ at 5 K and 1 T and 7 x $10^4$ and $10^5$ A/cm$^2$ at 20 K and zero field for samples A and B, respectively. $J_c$ for sample A is thus the same order as for sample B. This value is larger than for Nb sheathed $MgB_2$ tape which has a $J_c$ of 4.2 x $10^4$ A/cm$^2$ at 4.2 K and 1 T [12].

Although the $MgB_2$ thick films in our present work does not firmly attached to the steel substrate, it should be noted that the sintering conditions relating to cooling, heating rate and holding time still connot be considered optimum. The situation might be different in iron sheath $MgB_2$ tapes. We have found that the $MgB_2$ can strongly adhere to iron sheath materials. A thin layer of iron might be a suitable buffer layer between $MgB_2$ and other metal sheath materials which could react with $MgB_2$. For example, copper, which reacts with Mg and $MgB_2$ and form $MgCu_2$ at hight temperatures [20], could be coated with a very thin layer of iron. Composite of this type is under fabrication in our group.

## Conclusions

The $MgB_2$ thick films on stainless steel substrate can be fast formed by heating samples to 660 $^{o}C$ without any holding and quickly cooled down. XRD show above 90% $MgB_2$ phase. However, the samples sintered at 800 $^{o}C$ for 4 h contains $MgB_4$ and $MgB_2$ and MgO. The fast formed $MgB_2$ films appear to have a good grain connectivity that gives a $J_c$ of 8 x 10 $^{4}$ at 5 K and 1 T and maintained this value at 20 K in zero field. However, the $MgB_2$ films are not well adherent to steel substrate. The most possible reason may be big difference in thermal expansion coefficient between them. It is possible that a thin layer of ion can be used a buffer layer in other metallic sheath materials to $MgB_2$.

**Acknowledgments**

This work is partly support by the funding from Australian Research council and University of Wollongong.

**Figure captions**

Fig.1. Temperature profile for sample A (a) and sample (b).

Fig. 2. XRD for samples A (a) and B (b).

Fig 3. SEM surface images of samples A: (a) and (b) and B (c) and substrate (d). The magnification scale is the same in (c) and (d).

Fig.4. Temperature dependence of ac susceptibility of $MgB_2$ thick films for sample A (open circles) and B (solid line) after removal from the substrate.

Fig.5. Magnetisation hysteresis loops measured at 5 and 20 K. Open (5K) and closed circles (20K) represent sample A, and solid (5K) and dashed (20K) lines represent sample B.

Fig. 6. $J_c$ vs magnetic field. The symbols have the same meanings as in figure 4. The arrow shows the maximum field where the flux jumping ends.

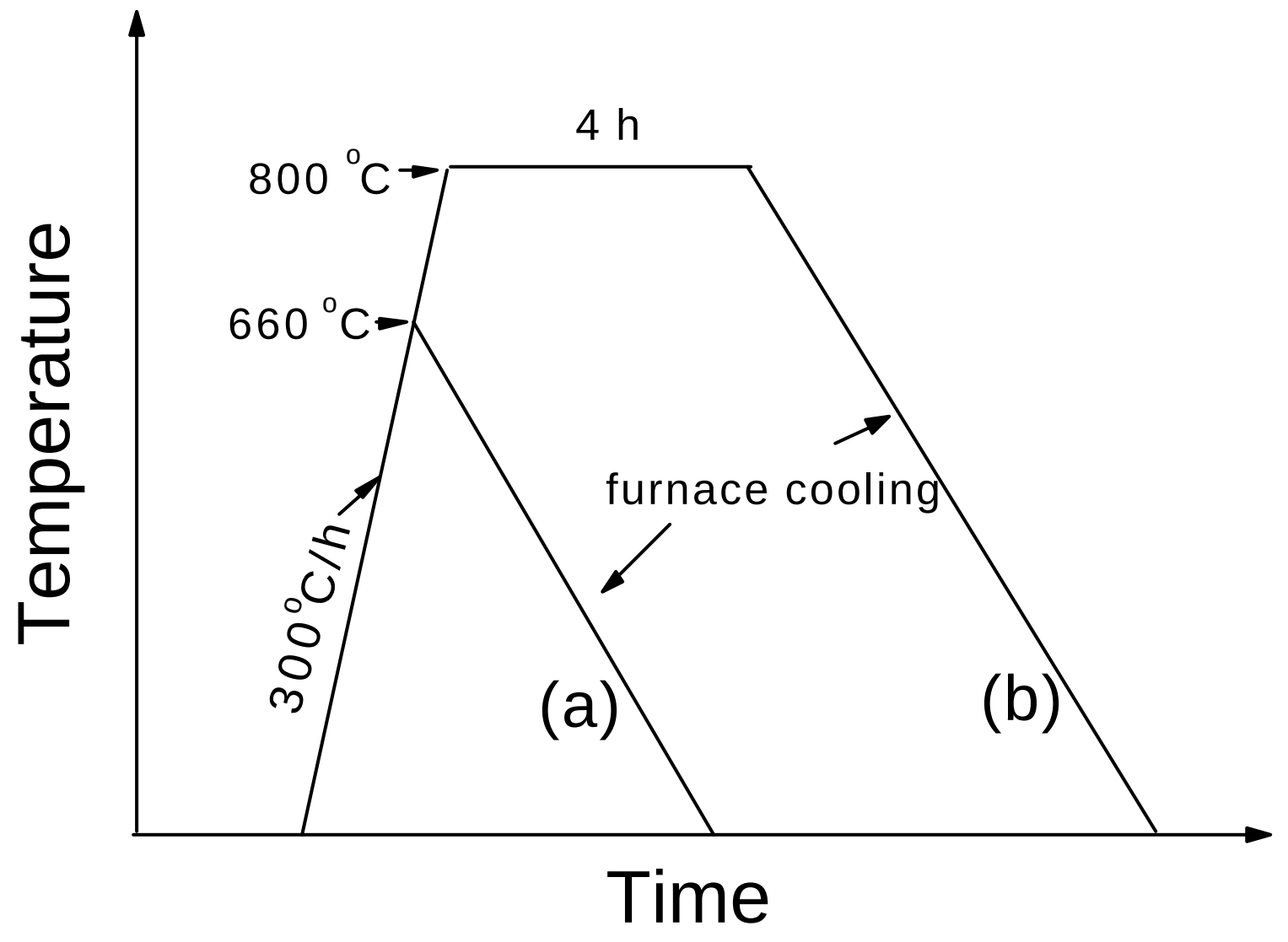

Fig. 1

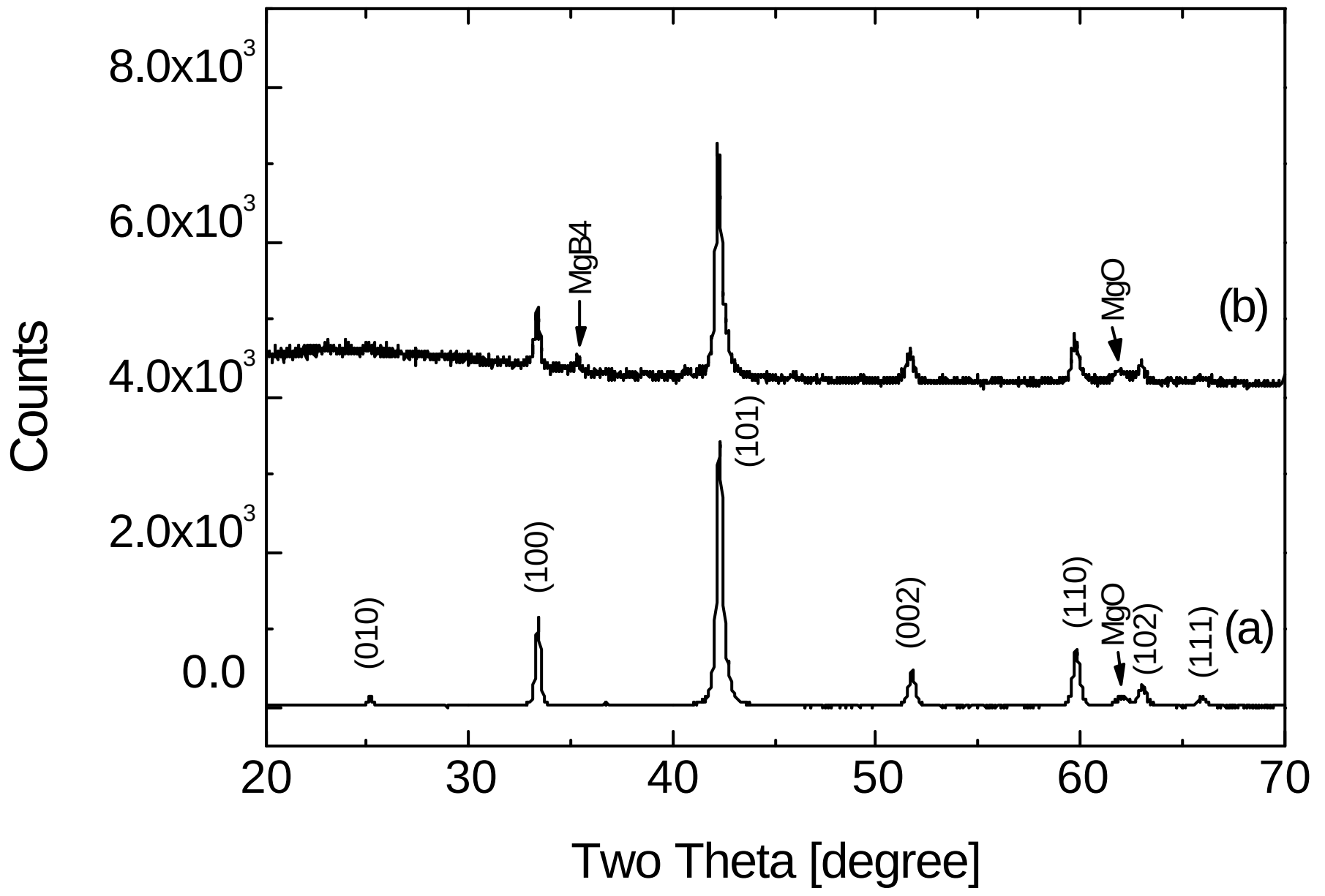

Fig. 2.

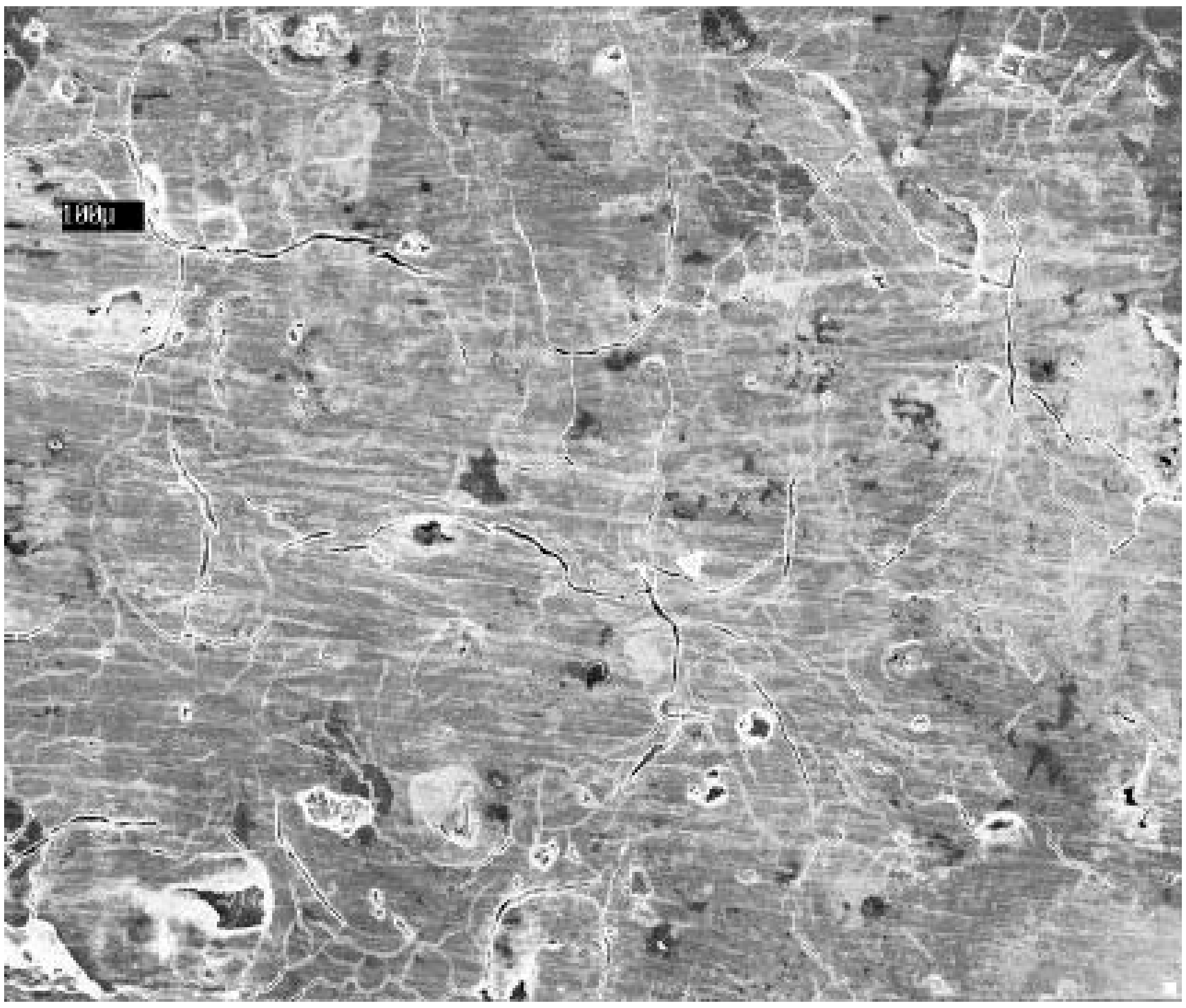

Fig. 3 (a)

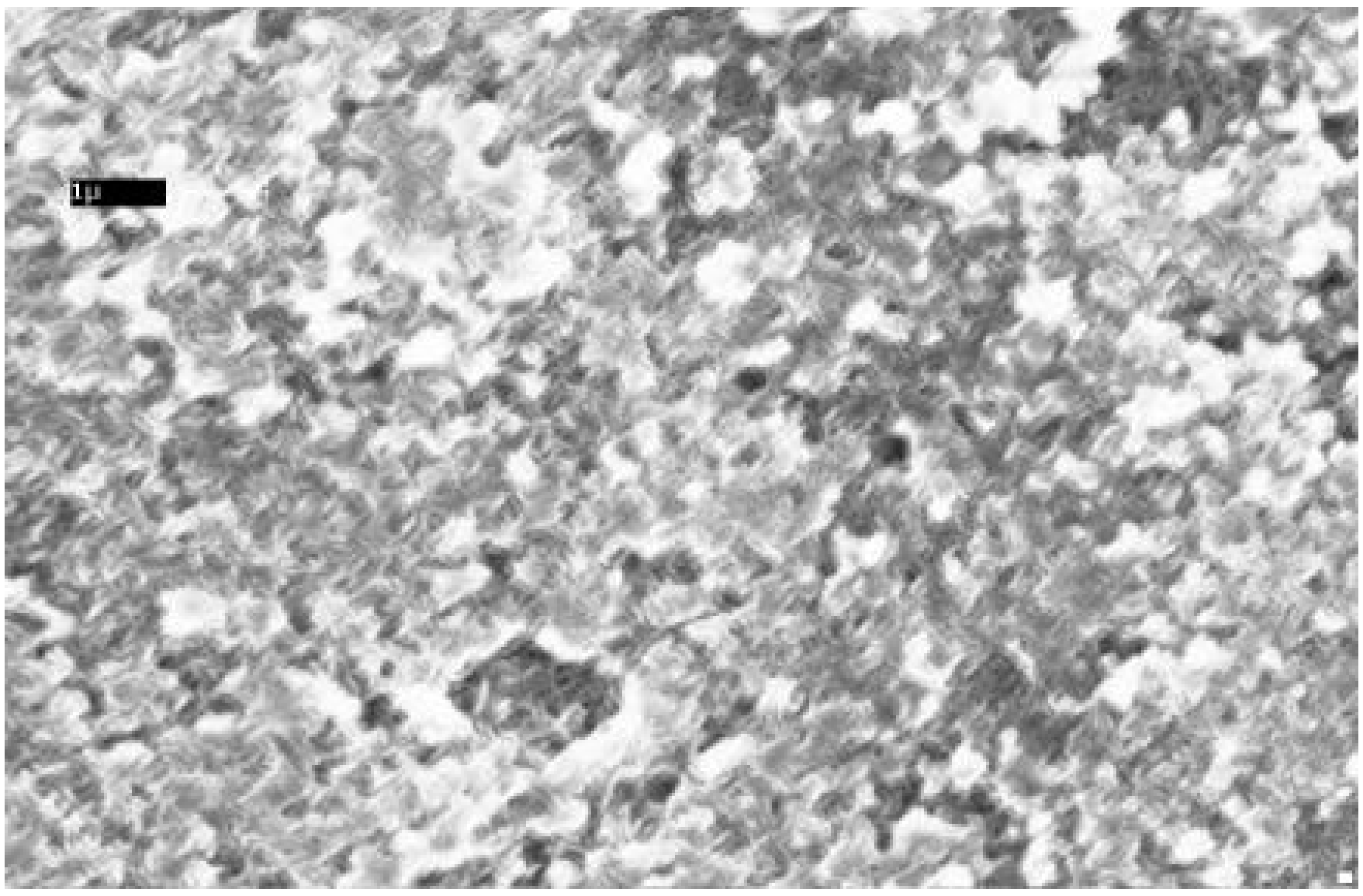

Fig. 3 (b).

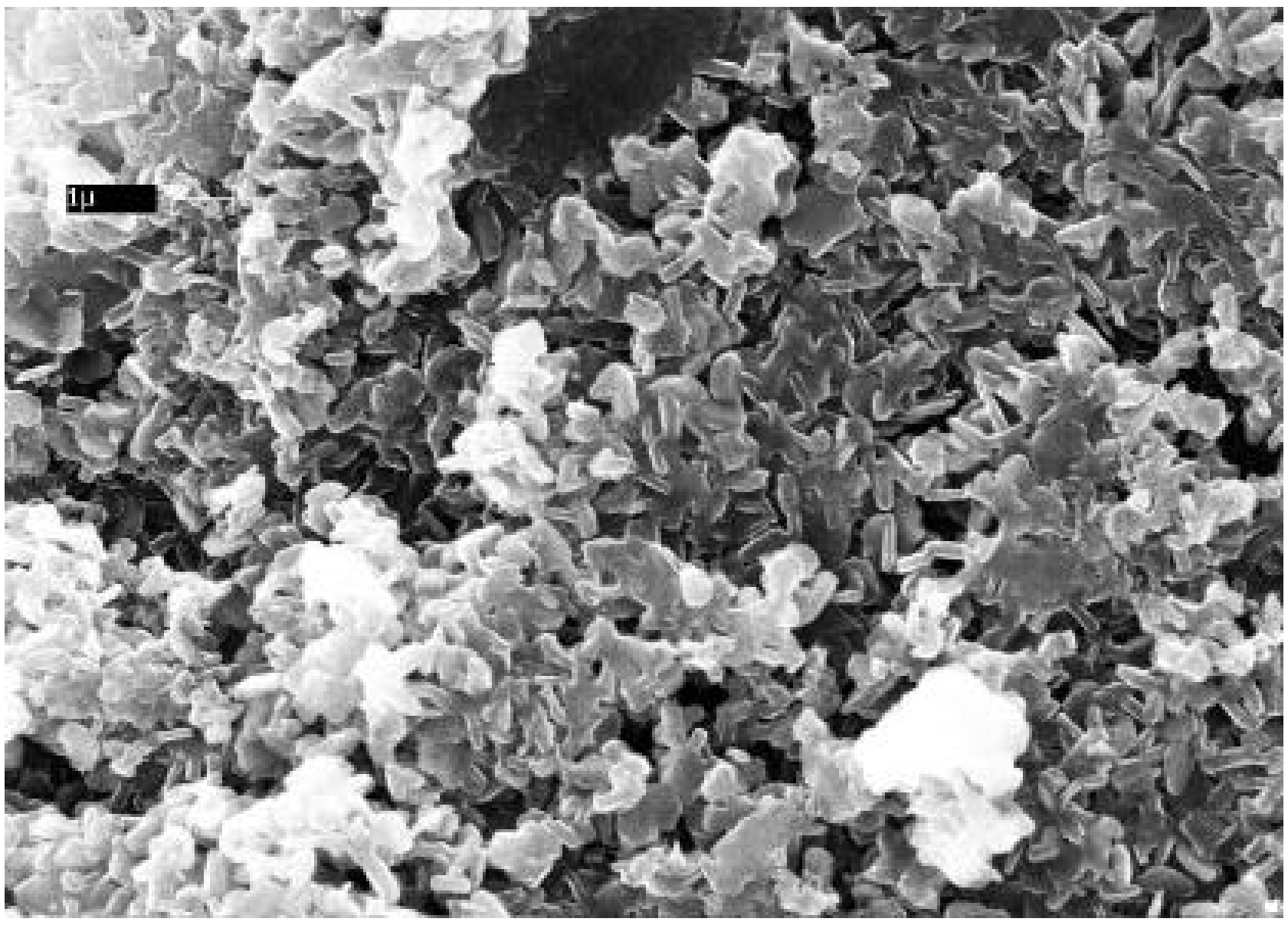

Fig. 3(c).

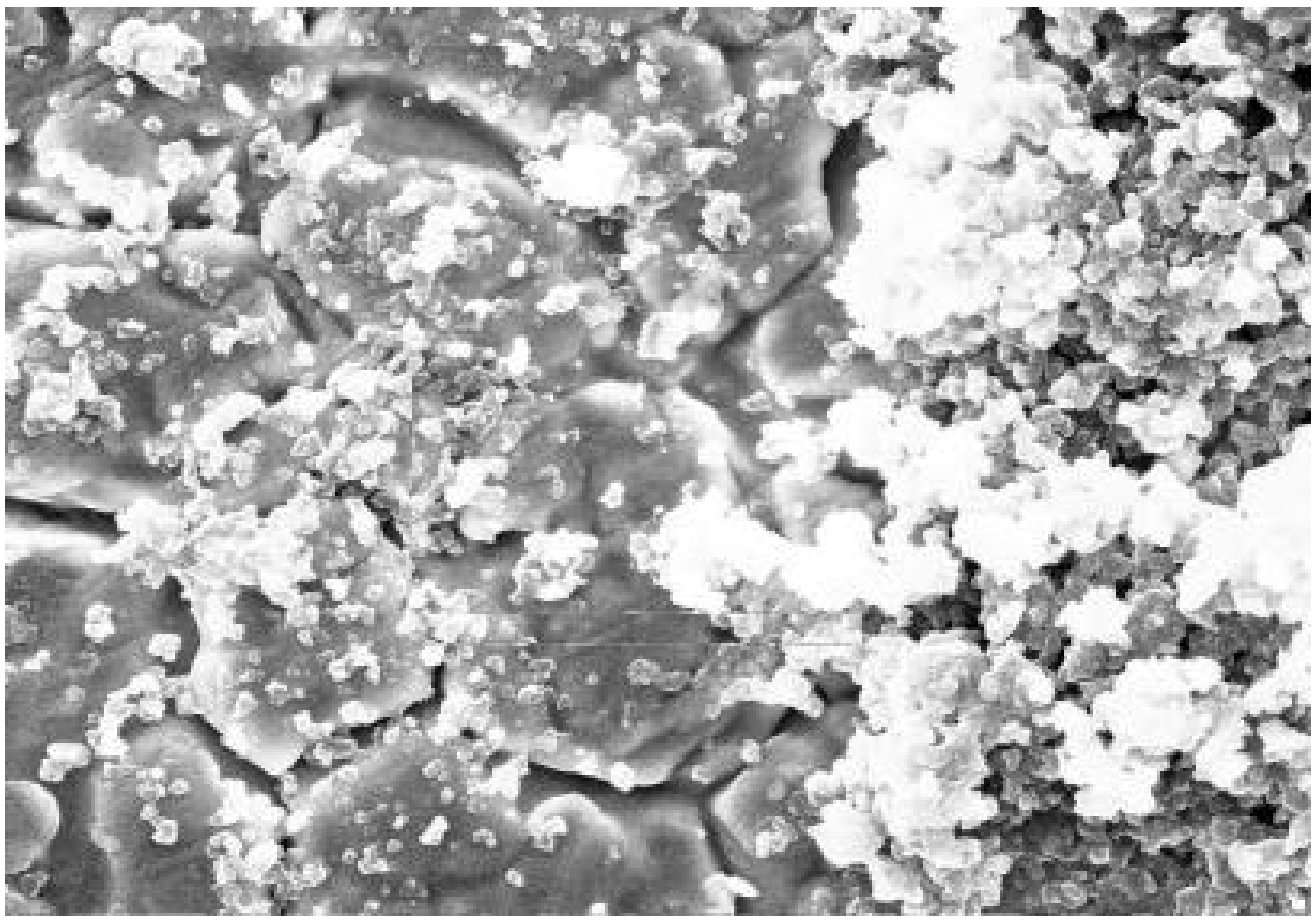

Fig. 3(d).

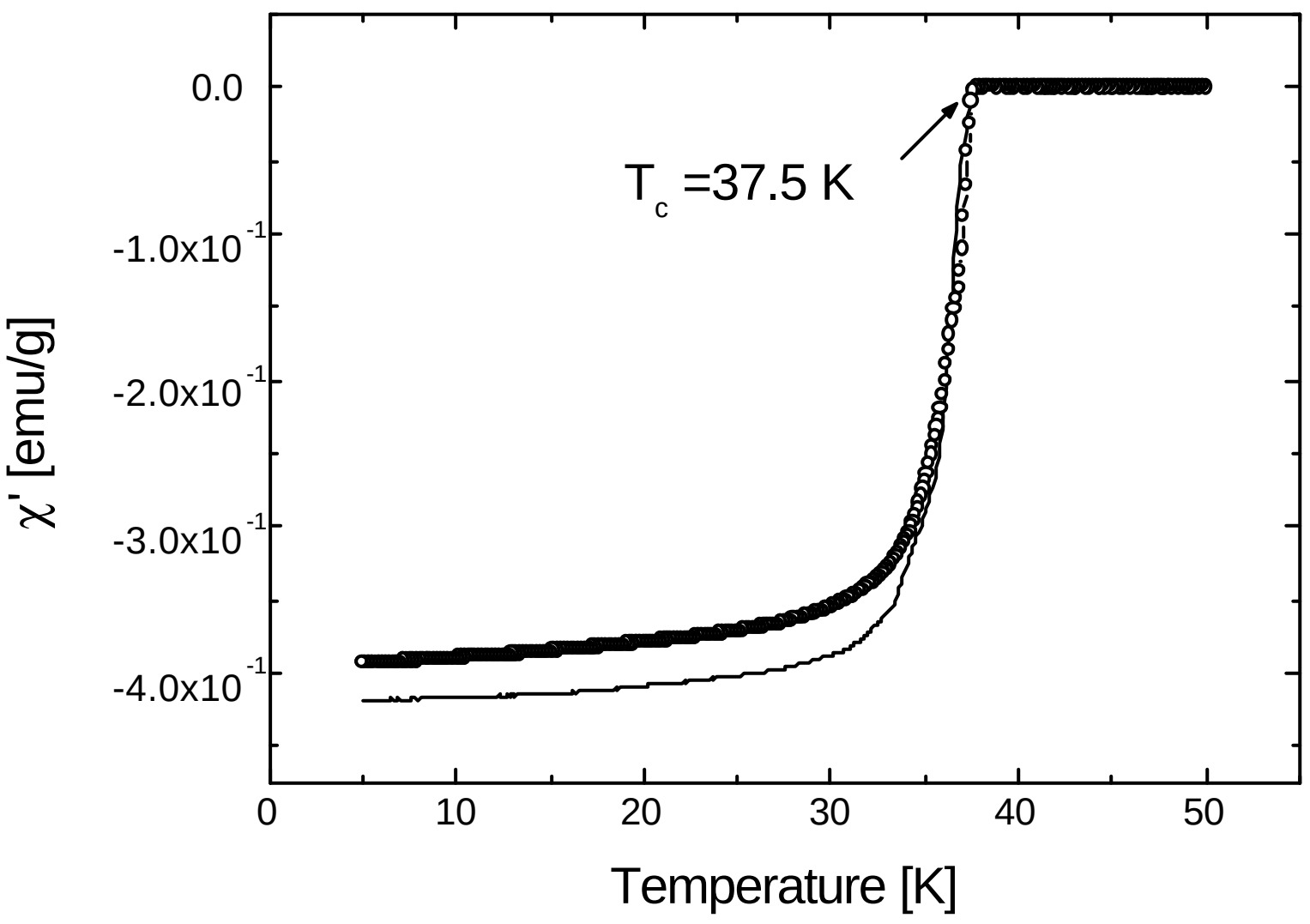

Fig. 4.

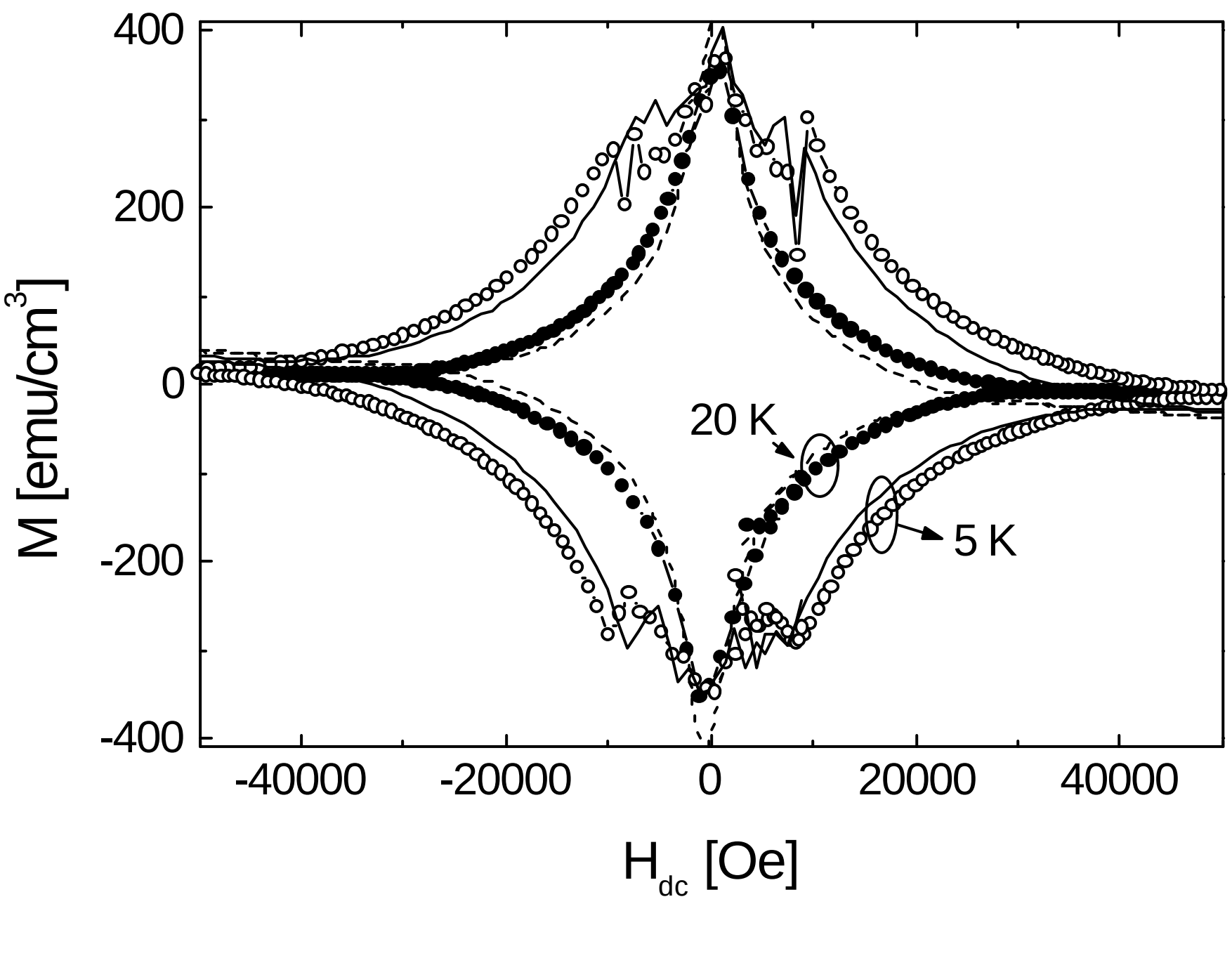

Fig. 5.

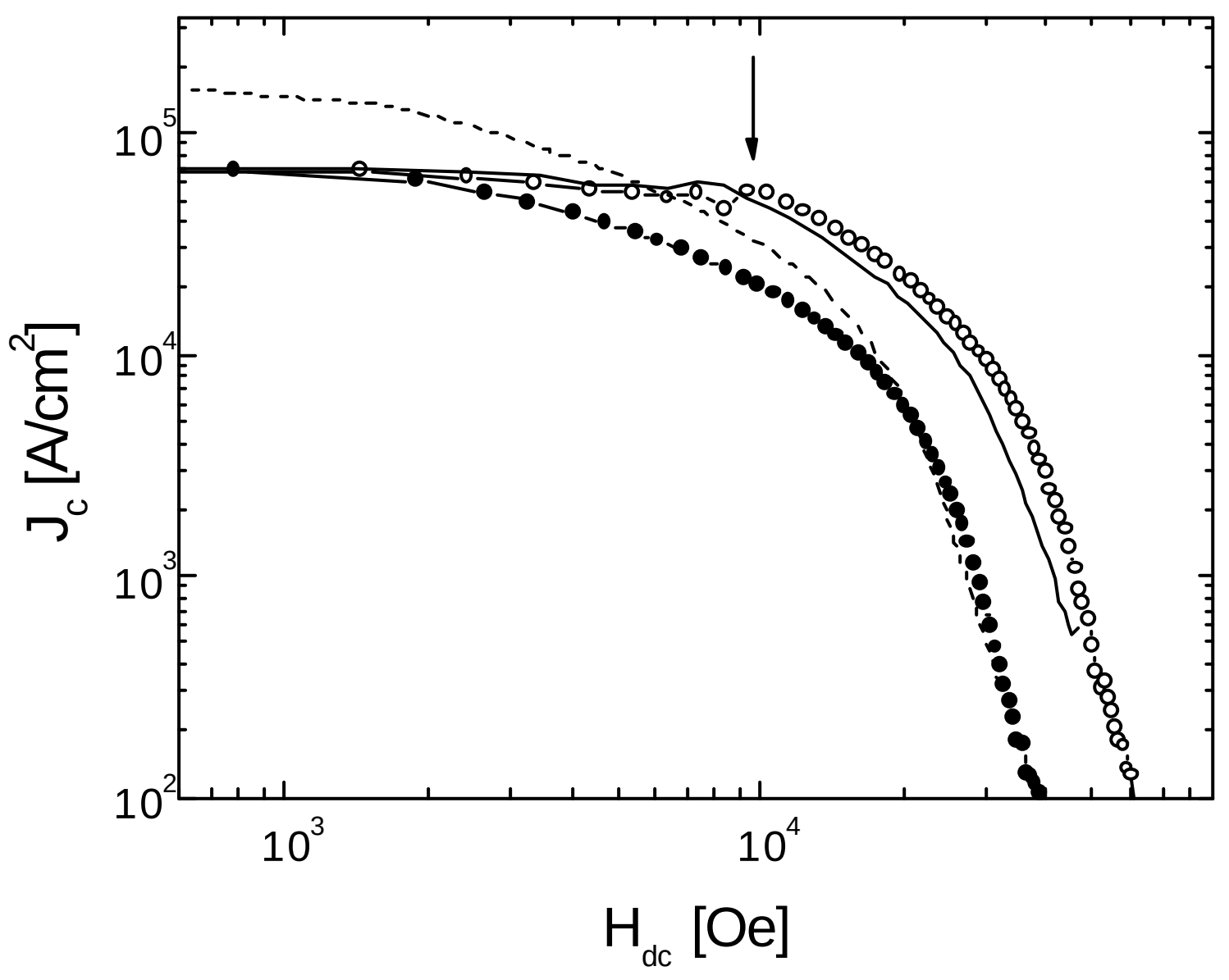

Fig. 6.